\documentstyle[12pt]{article}
 
\begin{document}

\baselineskip=14pt plus 0.2pt minus 0.2pt
\lineskip=14pt plus 0.2pt minus 0.2pt

\begin{flushright}
~~ \\ 
LA-UR-96-1597 \\
\end{flushright}

\begin{center}
\large{\bf Analytic Description 
of the Motion of a Trapped Ion in an Even or Odd Squeezed State} 

\vspace{0.25in}

\bigskip

 Michael Martin Nieto\footnote
{Email: mmn@pion.lanl.gov }

\vspace{0.3in}

{\it
Theoretical Division\\
Los Alamos National Laboratory\\
University of California\\
Los Alamos, New Mexico 87545, U.S.A.\\}

\vspace{0.3in}

{ABSTRACT}

\end{center}

\begin{quotation}

\baselineskip=0.333in

A completely analytic description 
is given of the motion of a trapped ion which is in either an even or 
an odd squeezed state.  Comparison is made to recent 
results on the even or
odd coherent states, and possible experimental work is discussed. \\

\vspace{0.25in}

\end{quotation}

\vspace{0.3in}

\newpage

\baselineskip=.33in

\section{Introduction}

The even ($+$) and odd ($-$) coherent states 
\cite{manko} can be defined as the eigenstates 
of the double-destruction operator, $aa$:
\begin{equation}
aa|\alpha\rangle_{\pm} = \alpha^2|\alpha\rangle_{\pm}~.
\end{equation}
They explicitly are
\begin{equation}
|\alpha\rangle_+ = [\cosh{|\alpha|^2}]^{-1/2}
   \sum_{n=0}^{\infty}\frac{\alpha^{2n}}{\sqrt{(2n)!}}|2n\rangle 
   \rightarrow \psi_{+}~,
\end{equation}
\begin{equation}
   |\alpha\rangle_- = [\sinh{|\alpha|^2}]^{-1/2}
   \sum_{n=0}^{\infty}\frac{\alpha^{2n+1}}{\sqrt{(2n+1)!}}|2n+1\rangle 
    \rightarrow \psi_{-} ~,
\end{equation}
where we will go back and forth between Dirac and wave-function notation.
These states also are the appropriate minimum-uncertainty 
coherent states. 

They can also be created by a special displacement operator
\cite{manko,knight}:
\begin{equation}
|\alpha\rangle_{\pm}= 
D_{\pm}(\alpha)|0\rangle
=\left[2(1\pm \exp[-2|\alpha|^2]\right]^{-1/2}
          \left[D(\alpha) \pm D(-\alpha)\right] |0\rangle~.
\end{equation}
where $D$ is the ordinary coherent state displacement operator:
\begin{equation}
D(\alpha)=\exp[\alpha a^{\dagger} - \alpha^* a]~,~~~~~~
\alpha = \alpha_1 + i \alpha_2 \equiv (x_0 + i p_0)/\sqrt{2}~.
\end{equation}
   
Matos Filho and Vogel \cite{vogel} 
have recently given a dynamical analysis, 
as a function of time, of a trapped ion
which, to very high precision, is in an even  or odd  coherent state. 
(Such a system has been produced experimentally by Wineland's group 
\cite{wineland}.)
They gave lovely three-dimensional numerical graphs  of the 
probability densities and Wigner functions, for the 
even and odd cases, as functions of position and time, for 
particular values of $\alpha$.

Previously, we had observed that closed-form expressions can be 
given for these wave functions in the time-independent 
case \cite{ntprl,ntpla}:
\begin{equation}
\psi_{+}=
\left[\frac{e^{-\alpha^2}}{\pi^{1/2}\cosh{|\alpha|^2}}\right]^{1/2}
e^{-x^2/2}\cosh(\sqrt{2}\alpha x)~,
\end{equation}
\begin{equation}
\psi_{-}=
\left[\frac{e^{-\alpha^2}}{\pi^{1/2}\sinh{|\alpha|^2}}\right]^{1/2}
e^{-x^2/2}\sinh(\sqrt{2}\alpha x)~.
\end{equation}
These expressions can be put in the form of  
two Gaussians displaced on opposite sides of the origin:  
\begin{equation}
\psi_{\pm}= 
\left[2 \pi^{1/2}(1 \pm e^{-2 |\alpha|^2})\right]^{-1/2} 
     \left[e^{-(x-\sqrt{2}\alpha_1)^2/2 + i\sqrt{2} \alpha_2 x} 
     \pm e^{-(x+\sqrt{2}\alpha_1)^2/2 - i\sqrt{2} \alpha_2 x}\right]~.
\end{equation}
where we have ignored  $e^{-i4\alpha_1 \alpha_2}$.  
This is an intuitively satisfying representation.

Then noting that time displacement 
can be included by letting $\alpha \rightarrow 
\alpha \exp[-i\omega t]$, we then could obtain an analytic 
expression for the wave functions
as a function of time \cite{eocs}.  Taking the convention 
$\alpha \rightarrow \alpha_0$ is real, as was done in 
Ref. \cite{vogel},  
the probability densities as a function of time were shown to be  
\begin{equation}
\rho_{+} =   \frac{e^{\alpha_0^2[\sin^2\omega t - \cos^2 \omega t]}}
              {\pi^{1/2} [e^{\alpha_0^2} +  e^{-\alpha_0^2}]}
           e^{-x^2}
         [\cosh\{2\sqrt{2}\alpha_0 (\cos\omega t)x\}
            + \cos\{2\sqrt{2}\alpha_0 (\sin\omega t)x\}],
\end{equation}
\begin{equation}
\rho_{-} =   \frac{e^{\alpha_0^2[\sin^2\omega t - \cos^2 \omega t]}}
              {\pi^{1/2} [e^{\alpha_0^2} -  e^{-\alpha_0^2}]}
           e^{-x^2}
         [\cosh\{2\sqrt{2}\alpha_0 (\cos\omega t)x\}
            - \cos\{2\sqrt{2}\alpha_0 (\sin\omega t)x\}].
\end{equation}
The Wigner functions can be obtained similarly.

The above  $\rho_{+}$ and $\rho_{+}$
described the forms of Figs. 1 and 4 in Ref. 
\cite{vogel} and   we show them in our Figs 1 and 2 (in time 
units of $\omega$, i.e. $\omega = 1$).  
The terms $\exp[-x^2] \times \cosh$  describe the two 
``wave-packets" on opposite sides of the origin.   Until they intersect,
these wave-packets resemble the non-spreading evolution of ordinary 
coherent states. The $\cos$  terms describe the 
interference effects near $x=0$ at $t=(2j+1)\pi/2$.   The even and odd 
natures are manifested by the maximum or zero at the origin,
respectively,  and the 
symmetry of the humps about the origin.  (Other discussions of  
``Schr\"odinger Cat" or ``two-packet" states should also be 
consulted \cite{wolfgang}.)

The question now arises if this formalism can be extended to squeezed
states. 


\section{Squeezed States}

For the even and odd systems, there is a well-defined 
mathematical prescription to obtain
ladder-operator and equivalent minimum-uncertainty squeezed states
\cite{ntprl}.  They are given, explicitly, as the eigenstates of the 
equation
\begin{equation}
\left[\left(\frac{1+q}{2}\right) aa
+\left(\frac{1-q}{2}\right) a^{\dagger} a^{\dagger} \right]\psi_{ss} 
=\alpha^2 \psi_{ss} .   \label{Kss}
\end{equation}
The solutions are  \cite{ntprl}
\begin{equation}
\psi_{Ess}=N_E\exp{\left[-\frac{x^2}{2}(q+\sqrt{q^2-1})\right]}
\Phi\left(\left[\frac{1}{4}+\frac{\alpha^2}{2\sqrt{q^2-1}}\right],~
\frac{1}{2};~x^2\sqrt{q^2-1}\right)~,
\end{equation}
\begin{equation}
\psi_{Oss}=N_O ~x\exp{\left[\frac{-x^2}{2}(q+\sqrt{q^2-1})\right]}
\Phi\left(\left[\frac{3}{4}+\frac{\alpha^2}{2\sqrt{q^2-1}}\right],~
\frac{3}{2};~x^2\sqrt{q^2-1}\right)~,
\end{equation}
where  $\Phi(a,b;c)$ is the confluent hypergeometric function
$ \sum_{n=0}^{\infty} \frac{(a)_n c^n}{(b)_n~n!}$.
In the limit $q\rightarrow 1$, these become the even and odd coherent states.

For the ordinary harmonic-oscillator 
coherent and squeezed states, there are equivalent 
displacement-operator squeezed states, since there exists a unitary 
Bogoliubov-type squeeze operator: 
\begin{equation}
S(z) = \exp\left[\frac{1}{2}za^{\dagger}a^{\dagger}-\frac{1}{2}z^*aa\right]~,
 ~~~~~~  z = re^{i\phi} = z_1 + i z_2~, 
\end{equation}
with the property
\begin{equation}
S^{\dagger}aS =(\cosh r) a + e^{i\phi} (\sinh r) a^{\dagger}~.
\end{equation}
In wave-function form, these states are \cite{njmp}
\begin{equation}
D(\alpha)S(z)|0\rangle = 
     \frac{\exp[-\frac{i}{2}x_0p_0]}{\pi^{1/4}[s(1+i2\kappa)]^{1/2}}
   \exp\left[-(x-x_0)^2\left(\frac{1}{2s^2(1+12\kappa)}-i\kappa\right)
      +ip_0x\right]~, \label{ss}
\end{equation}
\begin{equation}
s \equiv \cosh{r} + \frac{z_1}{r} \sinh{r}~, ~~~~~~~
\kappa \equiv \frac{z_2 \sinh r}{2rs}~.
\end{equation}
(For $z$ real, $\ln s = r$ sgn($r$).) 

But there are no equivalent 
displacement-operator squeezed states for the even/odd systems, because 
there is no unitary operator that can transform $aa$ into
the operator of Eq. (\ref{Kss}).  
However, an alternate idea is to 
simply use $S$ for the coherent even/odd systems, 
\begin{equation}
\psi_{s\pm} = D_{\pm}(\alpha)S(z)|0\rangle
\end{equation}
on the physical grounds 
that $S$ is the dilation operator \cite{kl}.  If one does that, then each
of the packets of the even/odd states will be of the form of Eq. (\ref{ss}).
Taking, for simplicity, the case $z$ real (or $\kappa = 0$), these states
are (ignoring an overall phase)
\begin{equation}
\psi_{s\pm} =\left[\pi^{1/2}2s(1\pm e^{-x_0^2/s^2-p_0^2s^2})\right]^{-1/2}
    \left[e^{-(x-x_0)^2/(2s^2)+ip_0x} \pm
          e^{-(x+x_0)^2/(2s^2)-ip_0x}\right]~.
\end{equation}

Now we compare $\psi_{E/Oss}$ with $\psi_{s\pm}$.  We do this in
Fig. 3.  There we compare $\rho_{Ess}=\psi_{Ess}^*\psi_{Ess}$, 
having parameters 
$\alpha = 2$ and $q=2$, with $\rho_{s+}=\psi_{s+}^*\psi_{s+}$, 
having parameters $\alpha = 
2$ (or $x_0 = 2\sqrt{2}$, $p_0 = 0$) and $s=3/2$.  These parameters were 
chosen not for the best overlap, but for a simple comparison of shapes.
One sees that the two probability densities  
are quite similar.   $\rho_{Oss}$ is even more similar to 
$\rho_{s-}$ because both are constrained to go to zero at the origin.  

Therefore, because of this similarity, and the analytic exactness of
the $\psi_{s\pm}$ system, we now proceed with this choice of squeezed 
states.  


\section{Time Evolution}

For the squeezed even/odd system, one no longer has the simple 
criterion that $\alpha \rightarrow \alpha e^{-i\omega t}$
describes the wave function.  

Instead we choose to consider the unitary time-evolution operator
(time again in units of $\omega$)
\begin{equation}
T = \exp[-iHt] = \exp[-i(a^{\dagger}a + 1/2)t]  
= \exp[-i(x^2 - \partial^2)/2]~.
\end{equation}
To make this operator useful, 
one can transform it to coordinates of the second 
kind with Baker-Campbell-Hausdorff relations.  (BCH relations are usually
obtained 
in terms of  raising and lowering operators, not in terms of 
the functional operators we have here.)  But when this is done, one 
obtains \cite{njmp}
\begin{equation}
T = [\cos t]^{-1/2} \exp[-\frac{i}{2} \tan t (x^2)]
\exp[-(\ln \cos t)(x\partial)]
\exp[\frac{i}{2} \tan t (\partial^2)]~,
\end{equation}
where the operational definitions on a function $h(x)$ are
\begin{eqnarray}
\exp[\tau(x\partial)] h(x) &=& h(xe^{\tau}) \\
\exp[c(\partial^2)] h(x) &=& 
\frac{1}{[4\pi c]^{1/2}} \int_{-\infty}^{\infty}
\exp\left[-\frac{(y-x)^2}{4c}\right] h(y) dy ~.
\end{eqnarray}

With this result, one can  calculate
\begin{equation}
\psi_{s\pm}(t) = U\psi_{s\pm}  ~.
\end{equation}
Taking, for simplicity, the case $z$ is real (or $\kappa= 0$) one has
\begin{eqnarray}
\psi_{s\pm}(t)
&=& \left[\frac{s}{2\pi^{1/2}(1\pm e^{-x_0^2 \cos^2 t})}
\frac{s^2 \cos t -i\sin t}{s^4 \cos^2 t + \sin^2 t}\right]^{1/2}
\nonumber \\
&~& \left\{\exp\left[-\frac{(x-x_0\cos t)^2}{2}
  \left(\frac{s^2-i\tan t}{s^4\cos^2 t + \sin^2 t}\right) 
    - \frac{i}{2}(\tan t)x^2\right] \right. \nonumber \\
&~& ~~~ \left.    
  \pm \exp\left[-\frac{(x+x_0\cos t)^2}{2}
  \left(\frac{s^2-i\tan t}{s^4\cos^2 t + \sin^2 t}\right) 
    - \frac{i}{2}(\tan t)x^2\right] \right\}~.
\end{eqnarray}
The terms $\exp[-i(\tan t)x^2/2]$ turn out to be necessary to cancel 
the singularities of the terms $\exp[ix^2\tan t/(2\sin^2t)]$ when 
$t$ is an odd multiple of $\pi/2$.  

Then some algebra yields 
\begin{eqnarray}
  \rho_{s\pm} &=& \frac{\exp[-(x^2+x_0^2\cos^2t)/d^2]}
                {\pi^{1/2}d[1\pm d\exp[-x_0^2/s^2]]}  \nonumber \\
       &~&~~\left\{\cosh\left(\frac{2xx_0(\cos t)}{d^2}\right)
            \pm \cos\left(\frac{2xx_0 \sin t}{d^2 s^2}\right)\right\}~,
\end{eqnarray}
where
\begin{equation}
d^2 = s^2 \cos^2t + \sin^2t/s^2~.
\end{equation}


\section{Discussion}

In Figures 4 and 5 we plot the probability densities $\rho_{s+}$ 
and $\rho_{s-}$, respectively, as functions of $x$ and $t$. 
This is done for parameters $x_0=4$ and $s = 2$.  This value of $s$ 
means the wave packets have a large $x$ uncertainty at $t=0$, when 
they are separated.  However, when they collide at the origin at 
$t=\pi/2$, their widths are narrow, and so the interference peak is 
much larger and much  more confined that was the case with the coherent
states ($s=1$.  Figure 6 has the same type of description, except that 
since this is an odd state, there is a null at the origin when $t=
\pi/2$, so that there are two smaller narrow peaks 
about the origin, which however are still much taller than the 
coherent-state peaks.

In Figures 6 and 7 we plot the probability densities $\rho_{s+}$ 
and $\rho_{s-}$, respectively, as functions of $x$ and $t$. 
This time it is done for parameters $x_0=4$ but $s = 1/2$. 
This value of $s$ 
means the separated wave packets  at $t=0$
have a small $x$ uncertainty.  Therefore,  when the 
packets collide at the origin at $t=\pi/2$, they have a large 
$x$ uncertainty, and so the interference pattern is very pronounced
and broad.  The evenness and oddness of the two figures is reflected 
by their being a single hump at the origin in Figure 6 and a null, 
surrounded symmetrically by humps,  in Figure 7.

The cases $s=2, 1/2$, are three examples in a continuum for $|z| = 2$. 
the first case has the phase $\phi=0$ and the second case has $\phi=\pi$. 
All other $\phi$ represent cases where the squeezing is rotated between 
the $s$ and $p$ phase-space coordinates, and so the maximum heights of the 
wave packets occur at times different than $t=0$ or $\pi/2$.  

Note that the $s>1$ case would have a strong experimental signal.  It would 
have a very strong signal at $ t= \pi/2$ that is very short in time.  
Wineland's group \cite{dw2} is independently trying to create such states. 


\section*{Acknowledgments}

I am happy to acknowledge the useful comments of 
Erwin Mayr, Wolfgang Schleich, 
Werner Vogel, and David Wineland.  This work was supported by the 
U. S. Department of Energy.  I also acknowledge the hospitality of the 
Abteilung f\"ur Quantenphysik,  University of Ulm, 
under auspices of the Alexander von Humboldt Stiftung, where this 
paper was completed.



\newpage

\baselineskip=.33in

\newpage
\large
\noindent{{\bf Figure Captions}}
\normalsize
\baselineskip=.33in

Figure 1.  A three-dimensional plot of the even-coherent-state probability density, $\rho_+$, as a function of position, $x$,
and time, $t$, for $\alpha_0 = 2$.

Figure 2.  A three-dimensional plot of the odd-coherent-state 
probability density, $\rho_-$, as a function of position, $x$,
and time, $t$, for $\alpha_0 = 5^{1/2}$. 

Figure 3. The dashed curve is a plot of   $\psi_{Ess}(x)$ vs. $x$,
with parameters 
$\alpha = 2$ and $q=2$. The normalization constant $N_E= 0.08190$. 
The solid curve is a plot of $\psi_{s+}(x)$ vs. $x$, 
with  parameters $\alpha = 
2$ (or $x_0 = 2\sqrt{2}$, $p_0 = 0$) and $s=3/2$.

Figure 4.  A three-dimensional plot of the even-squeezed-state 
probability density, $\rho_{s+}$, as a function of position, $x$,
and time, $t$, for $x_0 = 4$ and $s=2$.

Figure 5.  A three-dimensional plot of the odd-squeezed-state 
probability density, $\rho_{s+}$, as a function of position, $x$,
and time, $t$, for $x_0 = 4$ and $s=2$.

Figure 6.  A three-dimensional plot of the even-squeezed-state 
probability density, $\rho_{s+}$, as a function of position, $x$,
and time, $t$, for $x_0 = 4$ and $s=1/2$.

Figure 7.  A three-dimensional plot of the odd-squeezed-state 
probability density, $\rho_{s+}$, as a function of position, $x$,
and time, $t$, for $x_0 = 4$ and $s=1/2$.



\begin{thebibliography}{44}
\bibitem{kl} John Klauder has emphasized this point to me.
\end{thebibliography}
\end{document}